\documentclass[a4paper]{jpconf}
\usepackage{graphicx}
\usepackage{amsmath}
\usepackage{amssymb}

\newcommand{\be}{\begin{equation}}
\newcommand{\ee}{\end{equation}}
\newcommand{\bea}{\begin{eqnarray}}
\newcommand{\eea}{\end{eqnarray}}
\newcommand{\ei}{\end{itemize}}
\newcommand{\ben}{\begin{enumerate}}
\newcommand{\een}{\end{enumerate}}
\newcommand{\bt}{\begin{tabbing}}
\newcommand{\et}{\end{tabbing}}

\newcommand{\fKfpipm}{{f_{K^\pm}/f_{\pi^\pm}}}
\newcommand{\klth}{{K_{\ell_3}}}
\newcommand{\kltw}{{K_{\ell_2}}}
\newcommand{\piltw}{{\pi_{\ell_2}}}
\newcommand{\fpz}{{f_+(0)}}

\newcommand{\Vus}{{|V_{us}|}}
\newcommand{\Vud}{{|V_{ud}|}}

\begin{document}
\title{(Semi)leptonic kaon decays and neutral kaon mixing from lattice QCD}

\author{Takashi Kaneko}

\address{
  Theory Center, Institute of Particle and Nuclear Studies,
  High Energy Accelerator Research Organization(KEK), Ibaraki 305-0801, Japan
}
\address{
  School of High Energy Accelerator Science,
  The Graduate University for Advanced Studies (SOKENDAI), Ibaraki 305-0801, Japan
}
\address{
  Kobayashi-Maskawa Institute for the Origin of Particles and the Universe,
  Nagoya University, Aichi 464-8602, Japan  
}

\ead{takashi.kaneko@kek.jp}

\begin{abstract}
  We review recent progress on leptonic and semileptonic kaon decays
  and neutral kaon mixing from lattice QCD.
\end{abstract}


\section{Introduction}

Leptonic and semileptonic decays of kaons play a key role
in the determination of the Cabibbo-Kobayashi-Maskawa (CKM) matrix elements
$|V_{us}|$ and $|V_{ud}|$.
As reviewed by Flavor Lattice Averaging Group (FLAG)~\cite{FLAG21},
lattice QCD can predict relevant hadronic inputs,
namely the ratio of the kaon and pion decay constants $f_K/f_\pi$
and vector form factor at the zero momentum transfer $f_+(0)$,
at the sub-\% level
suggesting an intriguing unitarity violation,
the so-called ``Cabibbo angle anomaly''.
In this article,
as a member of FLAG,
we discuss a possible update of the latest FLAG review~\cite{FLAG21}
on the hadronic inputs and unitarity test.

Unfortunately,
there has not been much progress in the kaon mixing and, hence,
we briefly summarize the current status.


\section{Leptonic and semileptonic decays}

\begin{figure}[t]
  \includegraphics[width=0.47\linewidth]{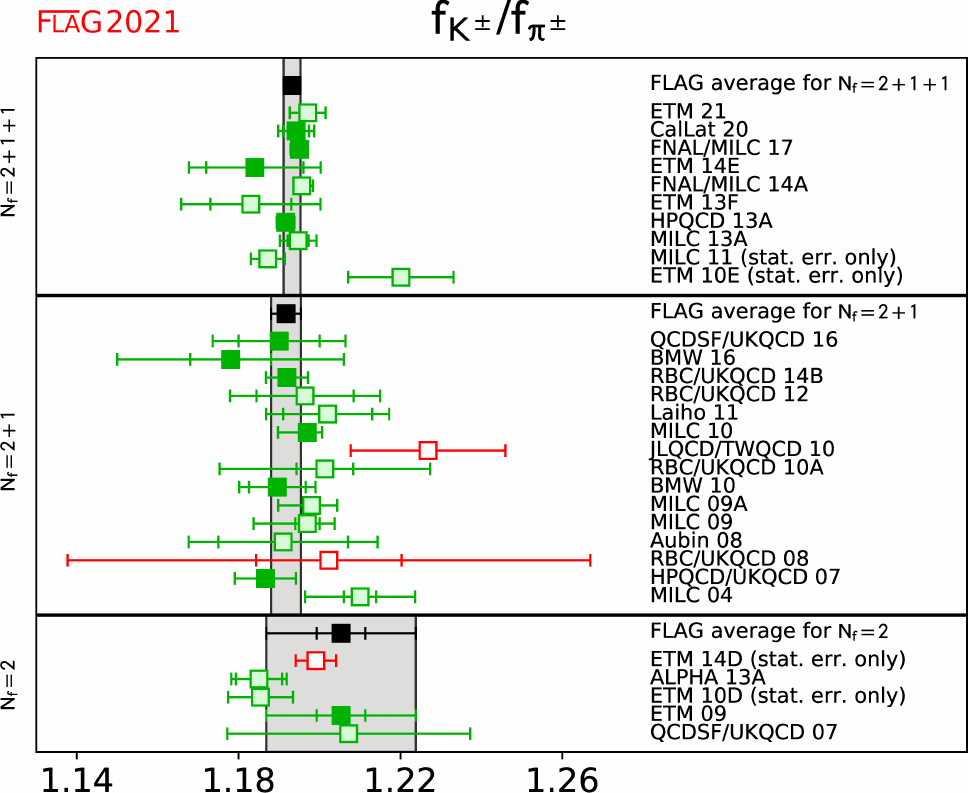}
  \hspace{3mm}
  \includegraphics[width=0.47\linewidth]{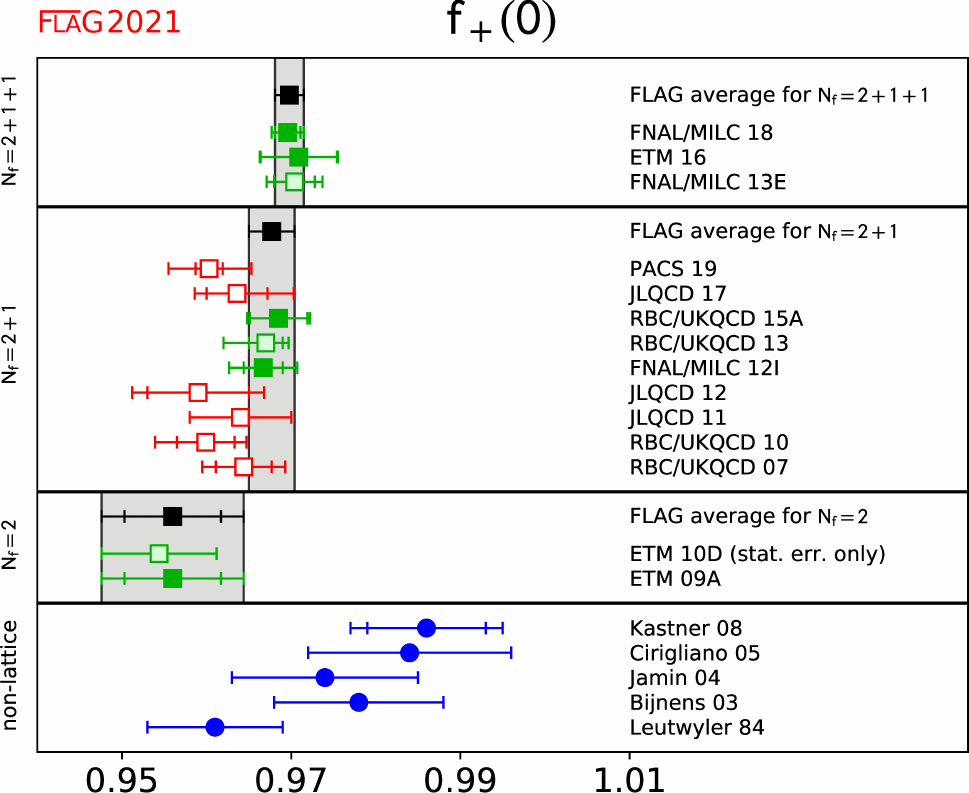}
  \vspace{2mm}
  \caption{
    \label{fig:flag21}
    Lattice results for $\fKfpipm$ (left panel) and $f_+(0)$ (right panel)
    discussed in the latest FLAG review~\cite{FLAG21}.
    Green squares are from studies satisfying FLAG's criteria
    for the control of systematic uncertainties,
    whereas simulation setup is not fully satisfactory for red squares.
    The black squares and bands show the FLAG average of
    the filled green squares for $N_f\!=\!2+1+1$, 2+1 and 2 QCD.
    The right panel also shows phenomenological estimate of $f_+(0)$
    (blue symbols).
  }
\end{figure}


Leptonic decays of kaons and pions, namely the $\kltw$ and $\piltw$ decays,
provide the determination of the ratio $\Vus/\Vud$ through
\bea
   \frac{\Gamma(K\to\ell\nu)}{\Gamma(\pi\to\ell\nu)}
   & = &
   \frac{\Vus^2}{\Vud^2}
   \left( \frac{f_K}{f_\pi} \right)^2
   \frac{M_K\left(1-m_\ell^2 / M_K^2\right)^2}
        {M_\pi\left(1-m_\ell^2 / M_\pi^2\right)^2}
   \left(1+\delta_{\rm EM}\right),
\eea
where $f_K$ and $f_\pi$ are kaon and pion decay constants in QCD,
respectively,
and the electromagnetic (EM) corrections to their ratio
is denoted by $\delta_{\rm EM}$.
A key advantage of this determination is that
systematic uncertainties,
such as the finite renormalization of the weak current on the lattice,
cancel (at least partially) in the ratio~\cite{VusVud}.


The left-panel of Fig.~\ref{fig:flag21} shows recent lattice QCD results
for $\fKfpipm$,
for most of which $\delta_{\rm EM}$ is estimated
in chiral perturbation theory (ChPT)
at next-to-leading order (NLO)~\cite{fKfpi:EM:chpt:1,fKfpi:EM:chpt:2}.
There have been many independent studies with good control of uncertainties.
The average quoted in the latest review is
$\fKfpipm\!=\!1.1932(21)$ for $N_f\!=\!2+1+1$  and
1.1917(37) for $N_f\!=\!2+1$ 
with 0.2\,--\,0.3~\% accuracy.


The ETM Collaboration recently carried out a precise independent calculation
in 2+1+1-flavor QCD at three lattice cutoffs $a^{-1}\!\lesssim\!2.9$~GeV
with the pion mass down to its physical value~\cite{fKfpi:Nf4:ETM}.
Three lattice volumes are simulated to control the finite volume effects.
The simulation setup is, therefore, satisfactory,
whereas it did not enter the latest FLAG average
simply due to its publication status.
Including this updates the average
as $\fKfpipm\!=\!1.1934(19)$ for $N_f\!=\!2+1+1$
with 0.16~\% accuracy,
whereas 1.1917(37) for $N_f\!=\!2+1$ remains unchanged.


With the precise hadronic input,
the uncertainty of $\delta_{\rm EM}\!\sim\!0.1$\,\% is no longer negligible.
While it is not easy to extend the ChPT estimate~\cite{fKfpi:EM:chpt:1,fKfpi:EM:chpt:2} to higher orders,
there has been recent progress in lattice QCD to calculate
the isospin corrections $\delta_{\rm iso}$ including $\delta_{\rm EM}$.
The Rome-Southampton group proposed
a sophisticated decomposition of the photon-inclusive decay rate
into infrared regular pieces, namely
the decay rate in the point-like approximation of the relevant mesons
and structure dependent correction
calculable on the lattice~\cite{fKfpi:EM:lat:Rome123:method}.
They obtain $\delta_{\rm iso}\!=\!-1.26(14)$\,\%~\cite{fKfpi:EM:lat:Rome123:result},
which is in nice agreement with the ChPT estimate $-1.12(21)$\,\%.
It is encouraging that an independent calculation
with a different lattice action by the RBC/UKQCD collaboration
obtained a consistent estimate~\cite{fKfpi:EM:lat:RBCUKQCD}.
We may expect more accurate estimate in the futute
by more realistic simulations and careful study of finite volume effects~\cite{fKfpi:EM:lat:RBCUKQCD:FVE}.



The $\klth$ semileptonic decays provide the determination of $\Vus$ through 
\bea
   \Gamma(K\!\to\!\pi\ell\nu)
   & = &
   \frac{G_F^2}{192\pi^3} \Vus^2 C_K^2 S_{\rm EW}
   M_K^5 I_{K\ell}
   f_+^{K^0\pi^-}(0)^2
   \left( 1 + \delta_{SU(2)}^{K\pi} + \delta_{\rm EM}^{K\ell} \right)^2,
\eea
where $G_F$ is the Fermi constant,
Clebsch-Gordan coefficient $C_K$ is 1 ($\sqrt{2}$)
for the neutral (charged) kason decay,
$I_{K\ell}$ is the phase-space integral.
The short distance electroweak correction is denoted by $S_{\rm EW}$,
whereas
$\delta_{\rm EM}^{K\ell}$ is the the long-distance EM correction.
Hadronic input is the vector form factor at zero-momentum transfer
$f_+^{K^0\pi}(0)$ for the reference channel $K^0\!\to\!\pi^-\ell\nu$,
which is simply denoted as $\fpz$ in the following.
And $\delta_{\rm SU(2)}^{K\pi}$ represents the strong isospin corrections
with respect to the reference channel.


The chiral expansion in ChPT can be written as $\fpz\!=\!1+f_2+\Delta f$,
where $f_2$ and $\Delta f$ represent the NLO and higher order corrections,
respectively. 
The Ademollo-Gatto theorem~\cite{AG} states that
these corrections are suppressed to $O((m_s-m_{ud})^2)$.
If we use $f_\pi$ instead of the decay constant in the chiral limit
in the chiral expansion,
the theorem also indicates that
there is no poorly known low-energy constants in $f_2$.
Therefore we can precisey determine $\fpz$ 
by calculating the small higher order correction $\Delta f$
with a reasonable accuracy on the lattice.

\begin{figure}[t]
  \includegraphics[width=0.6\linewidth,clip]{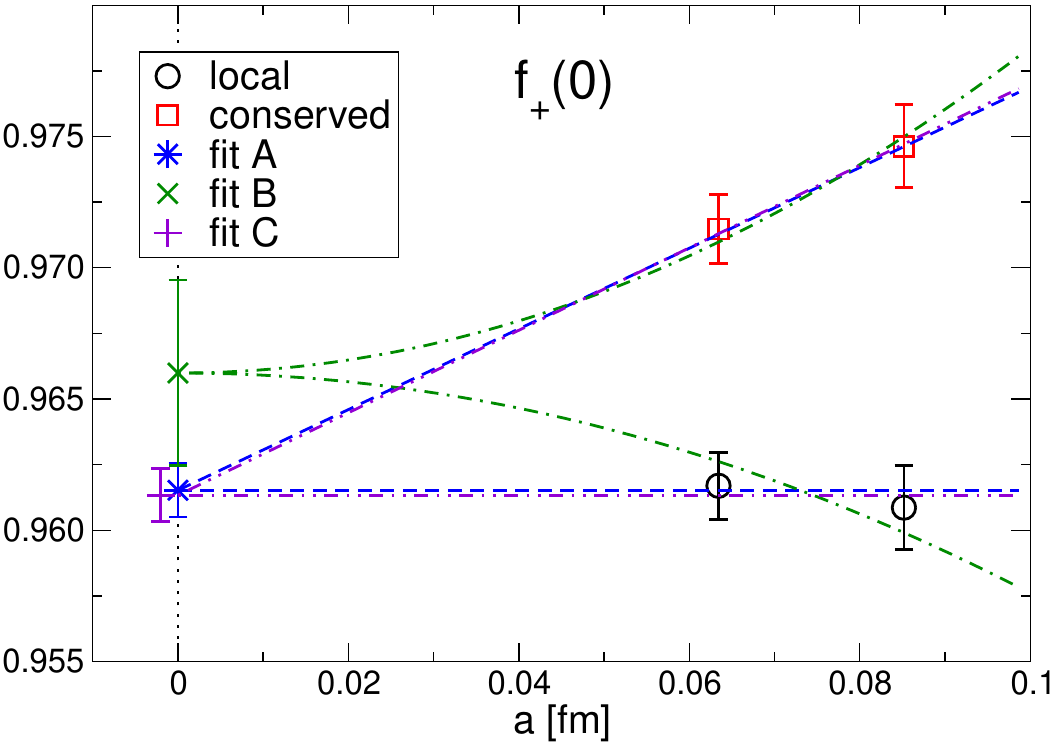}\hspace{2pc}%
  \begin{minipage}[b]{0.4\linewidth}
    \caption{
      \label{fig:f+0:pacs}
      Form factor $\fpz$ from Ref.~\cite{f+0:Nf3:PACS} as a function of $a$.
      The circles (squares) are obtained by using local (conserved)
      weak vector current on the lattice.
      They test three fitting forms of the lattice spacing and momentum transfer
      dependences (fit A, B and C), from which the dashed, dotted-dashed and
      dotted-dotted-dashed lines are reproduced.
      A larger value from fit B leads to a large and very asymmetric systematic
      error.
    }
    \vspace{10mm}
  \end{minipage} 
\end{figure}

The latest FLAG review covers lattice estimates of $\fpz$
shown in the right panel of Fig.~\ref{fig:flag21}
leading to the average $\fpz\!=\!0.9698(17)$ for $N_f\!=\!2+1+1$
and 0.9677(27) for $N_f\!=\!2+1$.
Recently, a new study for $N_f\!=\!2+1$
became available by the PACS collaboration~\cite{f+0:Nf3:PACS}.
Its characteric feature is that
they simulate the physical pion mass
on a large lattice volume of (10\,fm)$^3$
leading to good control of the chiral extrapolation
and finite volume effects.
It also enables them to simulate near-zero momentum transfers
with the standard periodic boundary condition.
With only two lattice spacings and unimproved current, however,
their largest and very asymmetric uncertainty comes
from the extrapolation to the continuum limit $a\!=\!0$
shown in Fig.~\ref{fig:f+0:pacs}.
Since their study does not fulfill the FLAG criterion,
which assumes the $O(a)$-improvement of the current,
the average remains unchanged for both $N_f\!=\!2+1+1$ and 2+1.
We note that 
PACS is simulating a smaller lattice spacing~\cite{f+0:Nf3:PACS:kaon22}
for a better control of the discretization effects.

With the 0.2\,\% accuracy of $f_+(0)$,
the uncertaintly of the EM corrections $\sim\!0.1$\,\% is not
negligible for the $\klth$ decays.
An extension of the Rome-Southampton method to semileptonic decays
is under investigation~\cite{f+0:EM:lat:Rome123}.
Another approach
based on ChPT with supplemented lattice data
has been proposed in Ref.~\cite{f+0:EM:lat+chpt:box}.


\section{CKM unitarity in the first row}

\begin{figure}[t]
  \includegraphics[width=0.6\linewidth,clip]{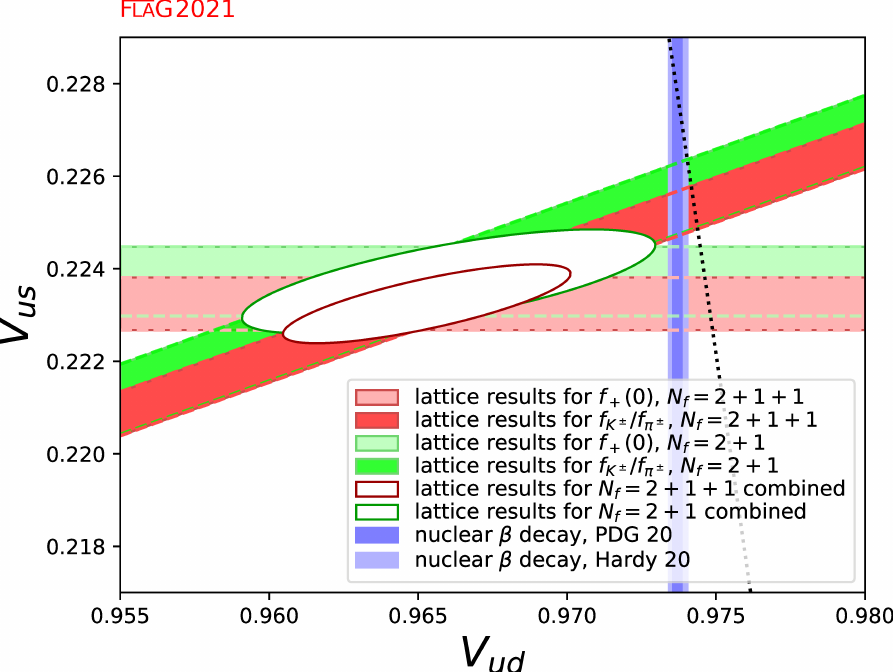}
  \hspace{0mm}
  \begin{minipage}[b]{0.4\linewidth}
    \caption{
      \label{fig:ckm}
      Test of CKM unitarity in $(|V_{ud}|,|V_{us}|)$ plane.
      The horizontal bands show $|V_{us}|$ determined from the $\klth$ decays,
      whereas $|V_{us}|/|V_{ud}|$ from the $\kltw$ and $\piltw$ decays
      is shown by the slanted bands. We plot results for $N_f\!=\!2+1+1$
      and 2+1 by the red and green bands, respectively.
      The ellipses show the $2\,\sigma$ contour of the intersection of
      the two bands. 
      We also plot $|V_{ud}|$ determined from the superallowed nuclear $\beta$
      decays by the vertical blue band. 
      The black dotted line satisfies unitarity in the first row.
    }
    \vspace{10mm}
  \end{minipage} 
\end{figure}


The latest FLAG review employs non-lattice inputs
$\Vus f_{K^\pm}/\Vud f_{\pi^\pm} \!=\! 0.2760(4)$ from PDG 20~\cite{PDG20}
and $\Vus f_+(0)\!=\!0.2165(4)$ from the CKM 2016 workshop~\cite{Vusf+0:CKM16:Moulson}\footnote{
There has been a slight update at CKM 2021~\cite{Vusf+0:CKM21:Moulson+Passemar}.
The change is well below $1\,\sigma$.}.
With this choice and hadronic inputs discussed above,
we obtain
\bea
   &&
   \frac{\Vus}{\Vud}
   =
   \left\{
   \begin{array}{ll}
     0.2313(5) & (N_f\!=\!2+1+1), \\
     0.2316(8) & (N_f\!=\!2+1),
   \end{array}
   \right.
   \hspace{5mm}
   \Vus
   = 
   \left\{
   \begin{array}{ll}
     0.2232(6) & (N_f\!=\!2+1+1), \\
     0.2237(7) & (N_f\!=\!2+1).
   \end{array}
   \right.
\eea        
These estimates are plotted in Fig.~\ref{fig:ckm}.
There is no significant devaiation between $N_f\!=\!2+1+1$ and 2+1,
and the former is slightly more precise with more recent studies.
Therefore, we focus on the results for $N_f\!=\!2+1+1$ in the following.


In order to test CKM unitarity in the first row, we calculate
$|V_u|^2 \!=\! \Vud^2 + \Vus^2 + |V_{ub}|^2$.
We note that
$|V_{ub}|$ is too small to have significant contribution to $|V_u|^2$,
and hence the long-standing tension between the exclusive and inclusive decays
is not problematic in the following test of unitarity.
With $\Vud/\Vus$ and $\Vus$ from kaon (semi)leptonic decays,
we obtain $|V_u|^2\!=\!0.9816(64)$ suggesting about $3\,\sigma$ tension
with unitarity.


The test sharpens considerably by combining the results from kaon decays
with the accurate estimate of $\Vud$ from superallowed nuclear $\beta$ decays.
In the following, we employ
$\Vud\!=\!0.97373(31)$~\cite{Vud:Hardy+Towner,PDG22},
which has been obtained with recent update of
the transition-independent~\cite{Vud:trans-indep:rad_corr}
and dependent~\cite{Vud:Hardy+Towner,Vud:trans-dep:rad_corr:Seng+,Vud:trans-dep:rad_corr:Gorchtein} radiative corrections.
With $\Vus$ from kaon semileptonic decay,
we obtain $|V_u|^2\!=\!0.99800(65)$,
which also shows $\sim\!3\,\sigma$ tension with unitarity
but much smaller uncetainty.
In contrast, 
$\Vus/\Vud$ from kaon leptonic decay leads to 
$|V_u|^2\!=\!0.99888(67)$, which is nicely consistent with unitarity.


Alternatively,
we can use the lattice estimate of the EM corrections $\delta_{\rm EM}$
to obtain $\Vus/\Vud\!=\!0.2320(5)$,
which is $1\,\sigma$ larger than that with $\delta_{\rm EM}$ from ChPT.
This result leads to better consitency among kaon leptonic decays,
superallowed nuclear decays and unitarity.
In contrast,
the tenstion between kaon (semi)leptonic decays and unitarity
is enhanced to 4~$\sigma$ with $|V_u|^2\!=\!0.9760(62)$.


\section{Neutral kaon mixing}

\begin{figure}[t]
  \includegraphics[width=0.6\linewidth]{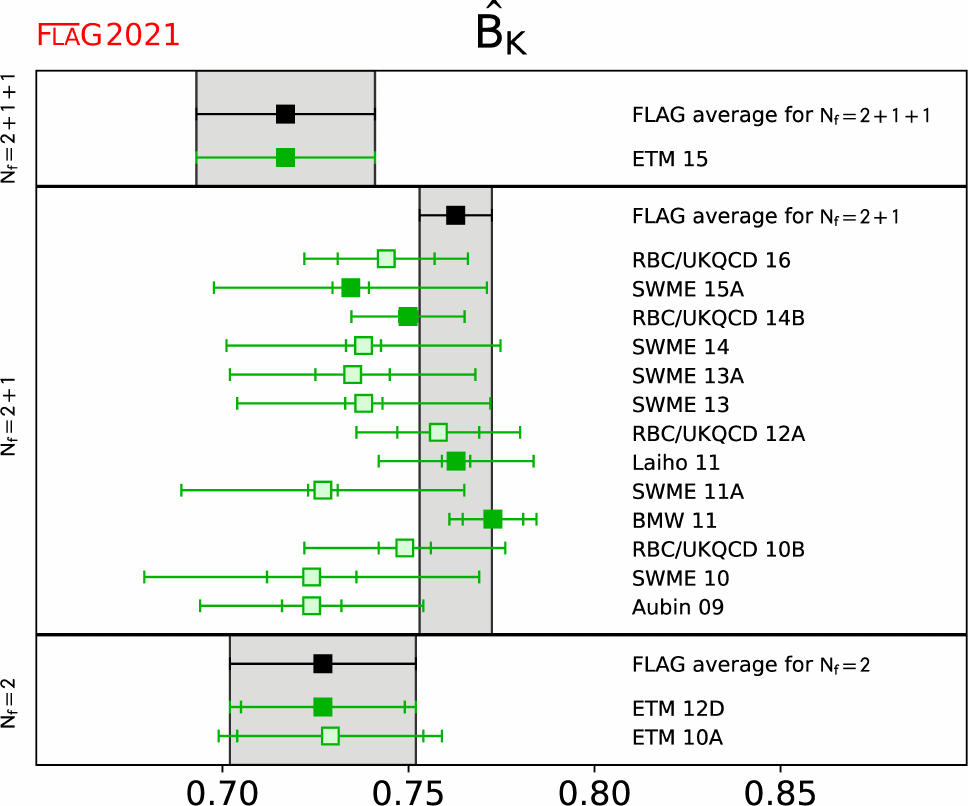}
  \hspace{0mm}
  \begin{minipage}[b]{0.4\linewidth}
    \caption{
      \label{fig:bk}
      Lattice results for bag parameter $B_K$~\cite{FLAG21}.
    }
    \vspace{10mm}
  \end{minipage} 
\end{figure}

The short distance contribution to the indirect CP violation parameter
$\epsilon_K$ is mediated by a local four-quark operator
${\mathcal O}_{\Delta s=2}\!=\!
\left\{\bar{s}\gamma_\mu\left(1-\gamma_5\right)d\right\}
\left\{\bar{s}\gamma_\mu\left(1-\gamma_5\right)d\right\}$.
Its matrix element is conventionally described by using the bag parameter
$B_K(\mu)$ defeined through
$(8/3) f_K^2 M_K^2 B_K(\mu) \!=\!
\left\langle \bar{K}^0 \left| {\mathcal O}_{\Delta s=2}
\right| K^0 \right\rangle$,
where $\mu$ represents the renormalization scale of 
the regularization scheme of one's choice.
Figure~\ref{fig:bk} shows lattice results for the renormalization group
independent bag parameter
$\hat{B}_K\!=\!\left(\bar{g}(\mu)^2 / 4\pi \right)^{-\gamma_0/(2\beta_0)}
{\rm exp}\left[ \int_0^{\bar{g}(\mu)} dg
\left( \gamma(g)/\beta(g) + \gamma_0/\beta_0 g \right) \right] B_K(\mu)$,
where
$\bar{g}(\mu)$ is the renormalized coupling,
$\beta(g)\!=\!-\beta_0 g^2/(4\pi)^2 - \cdots$ and 
$\gamma(g)\!=\!\gamma_0 g^2/(4\pi)^2 + \cdots$ are
the Callan-Symanzik $\beta$ function and anomalous dimension, respectively.
For $N_f\!=\!2+1$,
$\hat{B}_K$ has been calculated with the accuracy of about 1\,\% 
by using chiral symmetric quark action to avoid
unphysical operator mixing of ${\mathcal O}_{\Delta s=2}$
with other four-quark operators,
and/or by smearing gauge fields to suppress discretization effects. 
As a results,
the hadronic input $B_K$ is no longer source of the dominant uncertainty,
and there has not been much progress on $B_K$ in recent years.


In the effective Hamiltonian,
the CKM matrix element $|V_{cb}|$ appears in the Wilson coefficient
for ${\mathcal O}_{\Delta s=2}$,
and a domoinant uncertainty of the short disctance contribution 
arises from the long-standing tension in $|V_{cb}|$
between the exclusive and inclusive decays. 
To understand and revolve this tension,
there has been good progress in the lattice calculation of
the $B\!\to\!D^*\ell\nu$ form factors~\cite{review:lat22:B2DstarFF}
as well as the inclusive decay rates~\cite{review:lat22:heavy}.


With the current accuracy of $B_K$,
the uncertainty of the long-distance contribution is not negligible.
Note also that the $K_L$\,-\,$K_S$ mass difference $\Delta M_K$ is related with
the CP conserving part of the kaon mixing and thus long-distance dominated.
The lattice calculation of such long-distance contibutions
involves complicated correlation functions with two insertions of
the weak Hamiltonian, and hence is challenging.
However, there has been encouraging progress
by RBC/UKQCD collaboration~\cite{mixing:eK:long,mixing:dMK}.
The current lattice uncertainty of $\epsilon_K$ and $\Delta M_K$
is larger than experimental measurements,
and more realistic and independent simulations are highly welcome.

\section{Conclusion}

In this article, we review recent progress on (semi)leptonic decays
and mixing of kaons from lattice QCD.
There has been steady progress in the calculation of $\fKfpipm$ and $\fpz$,
the accuracy of which now reaches $\lesssim\,$0.2\,\%.
The latest estimate of $\Vus/\Vud$ and $\Vus$
from the kaon (semi)leptonic decays
suggests $3\,\sigma$ tension with CKM unitarity in the first row.
It is interesting to observe similar tension using
$\Vud$ from the superallowed nuclear $\beta$ decays
instead of $\Vus/\Vud$,
while the combination of the kaon leptonic and nuclear $\beta$ decays
show good agreement with unitarity.
However, we note that, as seen in Fig.~\ref{fig:flag21},
the average of $\fpz$ for $N_f\!=\!2+1+1$ is dominated by
the recent precise calculation by the Fermilab/MILC collaboration.
Independent and precise calculations are highly welcome to establish the tension
with unitarity.

At the $\lesssim\,$0.2\,\% accuracy of the hadronic inputs,
the uncertainty of the EM corrections is no longer negligible.
Lattice QCD is now providing the EM corrections for the leptonic decay,
which are consistent with the conventional ChPT estimate
and are systematically improvable.
Extension to the semileptonic decays is under development,
and would be also important for heavy meson decays,
whose accuracy is approaching to a few \% or less.

The accuracy of $B_K$ is already about 1\,\%.
To theoretically estimate $\epsilon_K$,
it is important to revolve the $|V_{cb}|$ tension
and to study the long distance contributions.
For the latter, the first exprolatory study has been already carried out,
but more realistic and independent studies are necessary.


\ack

We thank all members of Flavor Lattice Averaging Group,
in particular James N. Simone, Silvano Simula and Nazario Tantalo
for the $\Vus$ and $\Vud$ section, on which this review based.
We also thank Takeshi Yamazaki for helpful coresspondence.
The work of T.K. is supported in part by JSPS KAKENHI Grant Number
21H01085 and by the Post-K and Fugaku supercomputer project through the
Joint Institute for Computational Fundamental Science (JICFuS).
\vspace{5mm}


\end{document}